# Analysis of an SEIR-SEI four-strain epidemic dengue model with primary and secondary infections

*Raúl Isea*
Fundación Instituto de Estudios Avanzados
Hoyo de la Puerta
Baruta, República Bolivariana de Venezuela
*risea@idea.gob.ve*

***Abstract*— We propose an SEIR model for the populations and an SEI model for the vector to describe the transmission dynamics of a four-strain model with both primary and secondary dengue infections. In order to accomplish this, we propose and obtain an analytic solution of a system of 47 coupled differential equations. This would be the most complete epidemic model proposed to describe the dengue epidemic.**

## I. Introduction

Dengue is a viral disease, found in tropical and sub-tropical regions of the planet where it is estimated that between 2500 and 3000 million people are at risk of contracting the disease [1]. In fact, since 1998 the World Health Organization (WHO) has listed dengue as the tenth leading cause of death among all infectious diseases that are prevalent in the world [2].

Dengue is transmitted primarily by the bite of infected female mosquitoes *Aedes aegypti*, but it also has been associated with other species such as *Ae. albopictus*, *Ae. polynesiensis*, and *Ae. scutellaris* [3]. Dengue has four strains denoted by Dengue I-IV [1,2]. The most common form is the classic dengue or dengue fever that can often be caused by one strain. In addition, severe dengue hemorrhagic fever formerly associated with a secondary infection is caused by antibody-dependent enhancement processes [1,2].

The first case of dengue ocurred in Australia in 1954 and similar outbreaks were observed in the Philippines and subsequently spread to Vietnam, Thailand and other Asian countries [4]. The first epidemic of dengue in the Americas occurred in Cuba in 1981 which was caused by an Asian strain of Dengue serotype-II [5].

Due to the lack of effective drugs and vaccines against dengue fever there has been a lack of effective programs to help control the disease, and for this reason, a wide range of mathematical models to describe and characterize the dynamics of dengue transmission has been developed [6-8].

The classic example is the SIR model which indicates that there are three significant populations to be examined. They are the population that is susceptible to a given disease (S), the population that is infected with the disease (I) and the population that recovers from the disease (R). Aguilar *et. al.* [9] described an extension to numerically resolve dengue epidemics with four strains, employing a SIR model.

In the case of viruses, the mathematical model must include the incubation or latency period which occurs just before infection. In this case, the model is called the SEIR model where E represents the population that is exposed to the disease. In the case of dengue, the exposure time is approximately 8-9 days before manifestation of the disease once it is transmitted by an infected mosquito [10].

Our model proposes an analytic solution of a system of 47 differential equation that describes the dynamics of dengue transmission with four strains. In addition we take into account primary and secondary infections employing the SEIR model for the populations and the SEI model for vector; it is an extension of the model of Janreung and Chinviriyasit published in 2014 [11]. We believe this to be the most complete analytical analysis of the transmission of dengue.

.

## II. Mathematical model

The model is initially based on the recent model proposed by Janreung and Chinviriyasit [11] who resolved a system with 17 differential equations. In this model (Fig. 1), the host population (N) is subsequently subdivided into multiple populations based on the following assumptions:

-The model assumes a homogeneous mixture of the populations of both humans (host) and vectors (mosquito) so that each mosquito bite is as likely to transmit the virus to humans regardless of the type of the virus.

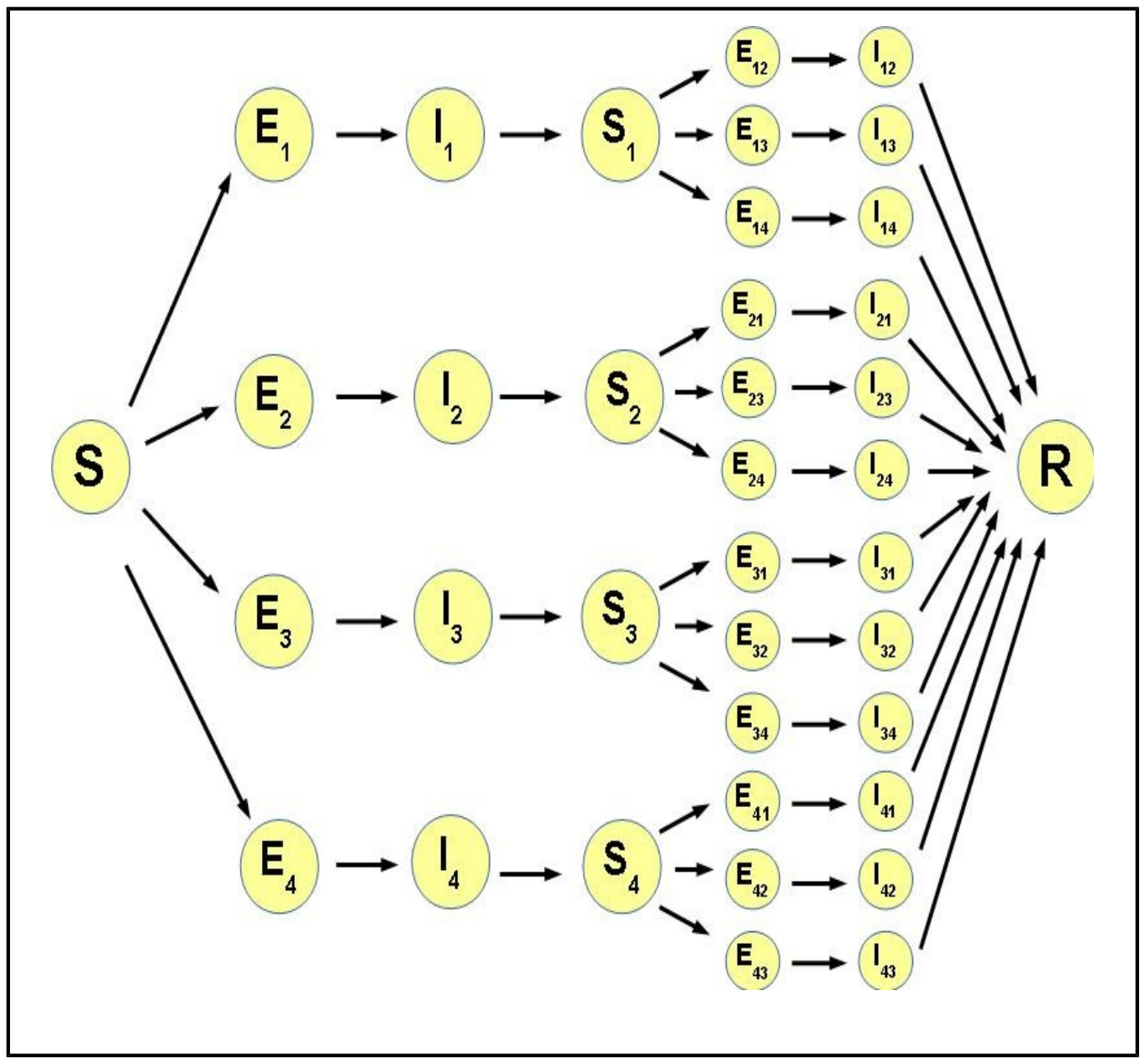

Fig. 1. The compartmental model of host-vector population employed in our model (see the text for the nomenclature).

-The total human population at time t is divided into 38 subpopulations. This means that the susceptible population will be exposed to an infection by one strain denoted by $E_{i.}$, where i= 1 to 4 is the index to indicate the four strains. Subsequently, the populations will become infected (denoted $I_i$). The secondary infection occurs when an individual is reinfected with another strain, and in this case, the population that is exposed to the second infection will be denoted as $E_{ij}$. The infected population will be identified as $I_{ij}$ (the second index represents the second infection while the first index represents the primary infection, with the condition i≠j). Finally the human population that recovered is indicated with R. The total population (N) will be equal:

$$N = S+E_1+E_2+E_3+E_4+I_1+I_2+I_3+I_4+S_1+S_2+S_3+S_4+ E_{12}+E_{13}+E_{14}+E_{21}+E_{23}+E_{24}+E_{31}+E_{32}+E_{34}+ E_{41}+E_{42}+E_{43}+I_{12}+I_{13}+I_{14}+I_{21}+I_{23}+I_{24}+I_{31}+I_{32}+I_{34}+ I_{41}+I_{42}+I_{43}+R$$

-The total mosquito (vector) population is denoted by M and is divided into 9 classes. The first is $S_v$ which represents the mosquito population susceptible to carry the virus. The next four are the mosquitoes that are exposed to the dengue virus and are denoted by $E_{vi}$. The last four, corresponding to infected mosquitoes, are denoted by $I_{vi}$. So

$$M = S_v+E_{v1}+E_{v2}+E_{v3}+E_{v4}+I_{v1}+I_{v2}+I_{v3}+I_{v4}$$

- It was also considered that the mosquitoes that were exposed to the four virus types at different times, once infected, cannot recover as has been already established in the scientific literature [9,11].

Finally, we model the system with a set of 47 ordinary differential equations which are written as:

$$\frac{dS}{dt}=\mu(N-S)-\sum_{i=1}^{4} S\frac{\beta}{N}I_{vi} \quad (1)$$

$$\frac{dE_i}{dt}=S\frac{\beta}{N}I_{vi}-(\sigma+\mu)E_i \quad (2)$$

$$\frac{dI_i}{dt}=\sigma E_i-(\gamma+\mu)I_i \quad (3)$$

$$\frac{dS_i}{dt}=\gamma I_i-S_i\frac{\beta}{N}\sum_{\substack{j=1\\ i\neq j}}^{4} I_{vj}-\mu S_i \quad (4)$$

$$\frac{dE_{ij}}{dt}=S_i\frac{\beta}{N}I_{vj}-(\sigma+\mu)E_{ij} \quad (5)$$

$$\frac{dI_{ij}}{dt}=\sigma E_{ij}-(\gamma+\mu)I_{ij} \quad (6)$$

$$\frac{dR}{dt}=\gamma\sum_{\substack{j=1\\ i\neq j}}^{4} I_{ij}-\mu R \quad (7)$$

$$\frac{dS_v}{dt}=M-S_v\frac{\beta_v}{N}\left(\sum_{i=1}^{4} I_i+\rho\sum_{\substack{j=1\\ i\neq j}}^{4} I_{ij}\right)-\mu S_v \quad (8)$$

$$\frac{dE_{vi}}{dt}=S_v\frac{\beta_v}{N}\left(I_i+\rho\sum_{\substack{j=1\\ i\neq j}}^{4} I_{ij}\right)-(\sigma_v+\mu_v)E_{vi} \quad (9)$$

$$\frac{dI_{vi}}{dt}=\sigma_v E_{vi}-(\gamma_v+\mu_v)I_{vi} \quad (10)$$

where i represent each strain with i= 1 to 4, and ρ is the rate of secondary infections contributing to intensity of the disease [12]. Of course, the best value of ρ is calculated according to the number of outbreaks of epidemics observed in each geographical area.

## III. Results

The epidemiologically relevant bioregion that is symbolized with Ω [9,11,12] is given by

$\Omega=\{ (S,E_1,E_2,E_3,E_4,I_1,I_2,I_3,I_4,S_1,S_2,S_3,S_4,E_{12},E_{13},E_{14}, E_{21},E_{23},E_{24},E_{31},E_{32},E_{34},E_{41},E_{42},E_{43},I_{12}, I_{13},I_{14}, I_{21},I_{23},I_{24},I_{31},I_{32},I_{34},I_{41},I_{42},I_{43},R,S_v,E_{v1},E_{v2},E_{v3},E_{v4}, I_{v1},I_{v2},I_{v3}, I_{v4}) \geq 0 \}$

The solution of this system of equations uses the same methodology as explained and published by Janreug and Chinviriyasit [11]. The solution of our system of equations suggests that there are two points of equilibrium that we call Point 1 and Point 2.

Point 1: This point correspond a disease free equilibrium analogous to that found in [11] equal to:

$(S'=N/\mu,0,0,0,0,0,0,0,0,0,0,0,0,0,0,0,0,0,0,0,0,0,0,0,0,0,0,0,0,0,0,0,0,0,0,0,0,0,0,0,0,0,0,S'_v=M/\mu_v,0,0,0,0,0,0,0,0)$
and,

Point 2: this is an endemic equilibrium point equal to:

$(S',E'_1,E'_2,E'_3,E'_4,I'_1,I'_2,I'_3,I'_4,S'_1,S'_2,S'_3,S'_4,E'_{12},E'_{13},E'_{14}, E'_{21},E'_{23},E'_{24},E'_{31},E'_{32},E'_{34},E'_{41},E'_{42},E'_{43},I'_{12}, I'_{13},I'_{14},I'_{21},I'_{23},I'_{24},I'_{31},I'_{32},I'_{34},I'_{41},I'_{42},I'_{43}, R',S'_v,E'_{v1},E'_{v2},E'_{v3},E'_{v4}, I'_{v1},I'_{v2},I'_{v3}, I'_{v4})$

The solution is found with the same mathematical methodology published in [11], and does contribute anything about new. The solution found is equal to:

$$S'=\frac{\mu N^2}{C_2} \tag{11}$$

$$E'_i=\frac{\mu\beta N I'_{vi}}{C_1 C_2} \tag{12}$$

$$I'_i=\frac{\beta\mu N\sigma I'_{vi}}{C_1 C_2 C_3} \tag{13}$$

$$S'_i=\frac{\beta\gamma\mu\sigma I'_{vi}}{C_1 C_2 C_3 D_i} \tag{14}$$

$$E'_{ij}=\frac{I'_{vi}\beta^2\gamma\mu\sigma I'_{vj}}{C_1^2 C_2 C_3 D_i} \tag{15}$$

$$I'_{ij}=\frac{I'_{vi}\beta^2\gamma\mu\sigma^2 I'_{vj}}{C_1^2 C_2 C_3^2 D_i} \tag{16}$$

$$S'_v=\frac{MN}{C_5} \tag{17}$$

$$E'_{vi}=\frac{M\beta_v}{C_5 C_6}\left[ I'_i+\rho\sum_{\substack{j=1\\ j\neq i}}^{4} I'_{ij}\right] \tag{18}$$

$$I'_{vi}=\frac{M\sigma_v\beta_v}{C_4 C_5 C_6}\left[ I'_i+\rho\sum_{\substack{j=1\\ j\neq i}}^{4} I'_{ij}\right] \tag{19}$$

In this case, the constants are:

$$C_1\equiv\sigma+\mu$$

$$C_2\equiv\mu N+\beta\left[\sum_{i=1}^{4} I'_{vi}\right]$$

$$C_3\equiv\mu+\gamma\,;\,C_4\equiv\gamma_v+\mu_v$$

$$C_5\equiv\mu_v N+\beta_v\left[\sum_{i=1}^{4} I'_i+\rho\sum_{\substack{j=1\\ j\neq i}}^{4} I'_{ij}\right]$$

$$C_6\equiv\sigma_v+\mu_v\,,$$

$$D_1\equiv\frac{\beta}{N}I'_{v2}+\frac{\beta}{N}I'_{v3}+\frac{\beta}{N}I'_{v4}+\mu$$

$$D_2\equiv\frac{\beta}{N}I'_{v1}+\frac{\beta}{N}I'_{v3}+\frac{\beta}{N}I'_{v4}+\mu$$

$$D_3\equiv\frac{\beta}{N}I'_{v1}+\frac{\beta}{N}I'_{v2}+\frac{\beta}{N}I'_{v4}+\mu$$

$$D_4\equiv\frac{\beta}{N}I'_{v1}+\frac{\beta}{N}I'_{v2}+\frac{\beta}{N}I'_{v3}+\mu$$

The most critical value in the epidemic model is the basic reproduction value (R0) and is the resulting higher value of the eigenvalues of the Jacobian of the 47 differential equations when evaluated for each critical point (*ie*., Point 1 and Point 2); the analysis is difficult to perform and will published in the future.

Finally, we will examine two cases of which the first results from the consideration of a single strain and the second results from the consideration of two strains.

Model 1. Single-strain model

In the case of a single strain i=1, so $S'_i=0, E'_{ij}=0, I'_{ij}=0.$ Therefore the solution is:

$$S'=\frac{\mu N^2}{\mu N+\beta I'_v}=\frac{N}{1+\frac{\beta I'_v}{\mu N}}$$

$$E_1^{'} \equiv E^{'} = \frac{\mu N \beta I_v^{'}}{(\sigma+\mu)(\mu N+\beta I_v^{'})}$$

$$I_1^{'} \equiv I^{'} = \frac{\beta \mu N \sigma I_v^{'}}{(\sigma+\mu)(\mu N+\beta I_v^{'})(\mu+\gamma)}$$

$$S_v^{'} = \frac{MN}{\mu_v N + \beta_v I^{'}}$$

$$E_{vi}^{'} \equiv E_v^{'} = \frac{M \beta_v}{(\mu_v N+\beta_v I^{'})(\sigma_v+\mu_v)} I^{'}$$

$$I_{vi}^{'} \equiv I_v^{'} = \frac{M \sigma_v \beta_v}{(\mu_v N+\beta_v I^{'})(\sigma_v+\mu_v)(\gamma_v+\mu_v)} I^{'}$$

Model 2. <u>Two-strains model</u>.

In this case, the values i are 1 and 2 and the terms that are not zero are S',E'$_1$,E'$_2$,I'$_1$,I'$_2$,,S'$_1$,S'$_2$,E'$_{12}$,E'$_{21}$,I'$_{12}$,I'$_{21}$,R',S'$_v$,E'$_{v1}$,E'$_{v2}$, I'$_{v1}$ and I'$_{v2}$ whose equations are respectively:

$$S^{'} = \frac{\mu N^2}{D_1^{'}}$$

$$E_1^{'} = \frac{\mu \beta N I_{v1}^{'}}{(\sigma+\mu)(D_1^{'})}$$

$$E_2^{'} = \frac{\mu \beta N I_{v2}^{'}}{(\sigma+\mu)(D_1^{'})}$$

$$I_1^{'} = \frac{\mu \sigma \beta N I_{v1}^{'}}{(\sigma+\mu)(\mu+\gamma)(D_1^{'})}$$

$$I_2^{'} = \frac{\mu \sigma \beta N I_{v2}^{'}}{(\sigma+\mu)(\mu+\gamma)(D_1^{'})}$$

$$S_1^{'} = \frac{\beta \gamma N^2 \mu \sigma I_{v1}^{'}}{(\sigma+\mu)(\mu N+\beta I_{v2}^{'})(\mu+\gamma) D_1^{'}}$$

$$S_2^{'} = \frac{\beta \gamma N^2 \mu \sigma I_{v2}^{'}}{(\sigma+\mu)(\mu N+\beta I_{v1})(\mu+\gamma) D_1^{'}}$$

$$E_{12}^{'} = \frac{I_{v1}^{'} \beta^2 \gamma N \mu \sigma I_{v2}^{'}}{(\sigma+\mu)^2(\mu N+\beta I_{v2}^{'})(\mu+\gamma) D_1^{'}}$$

$$E_{21}^{'} = \frac{I_{v2}^{'} \beta^2 \gamma N \mu \sigma I_{v1}^{'}}{(\sigma+\mu)^2(\mu N+\beta I_{v1}^{'})(\mu+\gamma) D_1^{'}}$$

$$I_{12}^{'} = \frac{I_{v1}^{'} \beta^2 \gamma N \mu \sigma^2 I_{v2}^{'}}{(\sigma+\mu)^2(\mu N+\beta I_{v2}^{'})(\mu+\gamma)^2 D_1^{'}}$$

$$I_{21}^{'} = \frac{I_{v2}^{'} \beta^2 \gamma N \mu \sigma^2 I_{v1}^{'}}{(\sigma+\mu)^2(\mu N+\beta I_{v1}^{'})(\mu+\gamma)^2 D_1^{'}}$$

$$S_v^{'} = \frac{MN}{D_3^{'}}$$

$$E_{v1}^{'} = \frac{M \beta_v (I_1+I_{12})}{D_3^{'} C_4}$$

$$E_{v2}^{'} = \frac{M \beta_v (I_2+I_{21})}{D_3^{'} C_4}$$

$$I_{v1}^{'} = \frac{M \sigma_v \beta_v (I_1+I_{12})}{D_3^{'} C_4 C_6}$$

$$I_{v2}^{'} = \frac{M \sigma_v \beta_v (I_2+I_{21})}{D_3^{'} C_4 C_6}$$

In this case, the constants are:

$$D_1^{'} \equiv (I_{v1}^{'}+I_{v2}^{'})\beta + \mu N$$

$$D_3^{'} \equiv \mu_v N + \beta_v (I_1+I_2+\rho(I_{12}+I_{21}))$$

The constants $C_4$ and $C_6$ keep their original definitions.

<u>Numerical simulations:</u>

In Table 1 lists the numerical values are listed of the parameters used in the simulation that were obtained from previous publication [11]. The time series plot is shown in Fig. 2(a) for the case of four strains for the time interval of 30 days. The solution with two strains is shown in Fig. 2(b). The results of both simulations are very similar.

## IV. Conclusions

We have developed an SEIR-SEI model of the transmission dynamics of four-strains of dengue model considering primary and secondary infections, which is an extension of the previous model proposed in the literature but only only considering two strains. We found that this model has two equilibrium points: the disease free equilibrium (called Point 1) and Point 2 which is the endemic equilibrium of the system. It is interesting to indicate that the model that analyzing the epidemic with four strains is very similar with to the model with two strains (Fig. 2), and therefore we conclude that it is unnecessary to perform analytical studies with four strains of dengue, since it is sufficient to consider a solution with two different strains.

## Acknowledgment

The author wishes to express his sincere thanks to Prof. Karl Lonngren and Johan Hoebeke for their unconditional help and the comments concerning the manuscript.

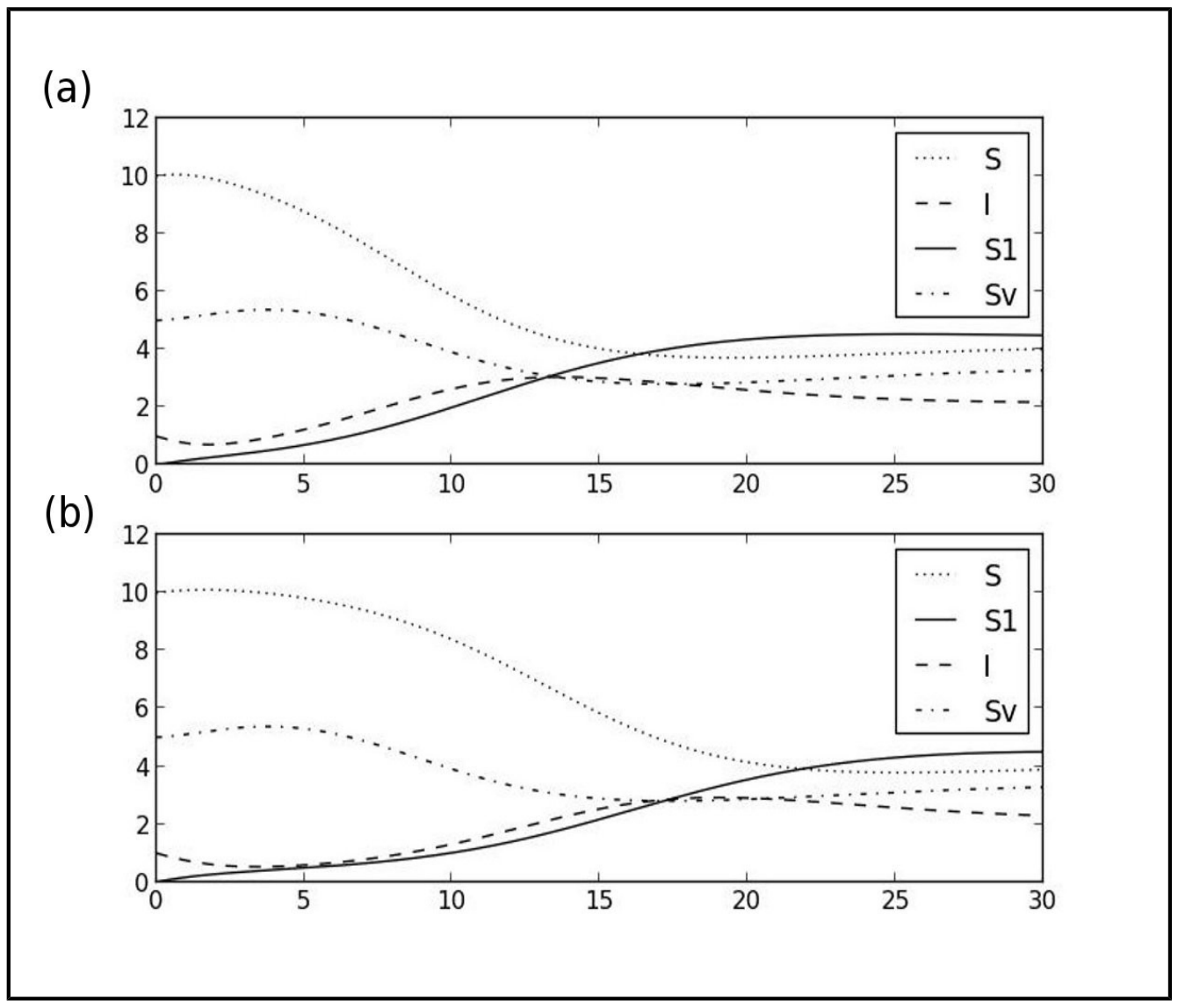

Fig. 2. The numerical solution of this model with (a) four strains, and (b) two strains according to Table 1 values. S represent the susceptible population, I is equal to the sum of $I_1+I_2+I_3+I_4$. $S_1$ is the humans exposed with strain 1, and finally $S_v$ is the total population of mosquitoes susceptible to any strains.

## References

[1] I. Kurane and T. Takasaki, "Dengue fever and dengue hemorrhagic fever: challenges of controlling an enemy still at large," Rev. Med. Virol. Vol. 11, pp. 301-311 (2001).

[2] A. Seijo, "El dengue como problema de salud pública," Arch. Argent Pediatr., vol. 99, pp. 510-521 (2001).

[3] L. A. Hill, J. B. Davis, G. Hapgood, P. I. Whelan, G. A. Smith, S. A. Ritchie, R. D. Cooper, and A.F. van den Hurk, "Rapid Identification of Aedes albopictus, Aedes scutellaris, and Aedes aegypti Life Stages Using Real-time Polymerase Chain Reaction Assays," Am. J. Trop. Med. Hyg., vol. 79(6), pp. 866–875 (2008).

[4] S.B. Halstead, "The XXth Century dengue pandemic: need for surveillance and research," World Health Stat Q, vol. 45, pp. 292-298 (1992).

[5] G. Kourí, M.G. Guzmán, J. Bravo J, "Hemorrhagic dengue in Cuba: history of an epidemic," Bulletin of the Pan American Health Organization, vol. 20, pp. 24-30 (1986).

[6] H. S. Rodrigues, M. T. Monteiro, D. F. Torres, "Vaccination models and optimal control strategies to dengue," Math Biosci, vol. 247, pp. 1-12 (2014).

[7] G. Chowell, R. Fuentes, A. Olea, X. Aguilera, H. Nesse, J. M. Hyman, "The basic reproduction number R0 and effectiveness of reactive interventions during dengue epidemics: the 2002 dengue outbreak in Easter Island, Chile," Math Biosci Eng., vol. 10, pp. 1455-1474 (2013).

[8] M. Canals, C. González, A. Canals, D. Figueroa, "Dinámica epidemiológica del dengue en Isla de Pascua," Rev. chil. Infectol., vol.29, pp. 388-394 (2012).

[9] M. Aguilar, B.W. Kooic, F. Rochaa, P. Ghaffari, N. Stollenwerk, "How much complexity is needed to describe the fluctuations observed in dengue hemorrhagic fever incidence data?," Ecological Complexity, vol. 16, pp. 31–40 (2013).

[10] M. Chan, M.A. Johansson, "The Incubation Periods of Dengue Viruses," PLoS ONE, vol. 7(11), pp. E50972 (2012).

[11] S. Janreung and W. Chinviriyasit, "Dengue Fever with Two Strains in Thailand," IJAPM., vol. 4, pp. 55-61 (2014).

[12] A. Korobeinikov, "Global Properties of SIR and SEIR Epidemic Models with Multiple Parallel Infectious Stages," Bull Math Biol., vol. 71, pp. 75–83 (2009).

Table 1. Parameters of the system of differential equations for the model SEIR-SEI epidemic dengue model

| Variables | Description | Value |
|---|---|---|
| S | Population of humans susceptible of any strains | S(0) = 10 |
| β | infection rate of the disease in population | 0.9 |
| βv | infection rate of the disease in vector | 1.0 |
| Si | Humans susceptible with strain i | $S_1 = 1, S_2=S_3=S_4=0$ |
| Ei | Humans exposed with i strains | All values are zero |
| Ii | Humans infected with strain i | $I_1 = 5, I_2=I_3=I_4=0$ |
| Eij | Humans infected with strain i but susceptible to strain j | All values are zero |
| Iij | Humans infected with strain i and reinfected with j strains | All values are zero |
| Sij | Humans susceptible of strain j but infected with strain i | All values are zero |
| Sv | Population of mosquitoes susceptible to any strains | 1 |
| Evi | Mosquitoes exposed with strain i | All values are zero |
| Ivi | Mosquitoes infected with strain i | All values are zero |